# PHz scale spectral broadening and few-cycle compression of Yb:KGW laser pulses in a pressurized, gas-filled hollow-core fiber

Z. Pi, H. Y. Kim, and E. Goulielmakis*

*Institute of Physics, University or Rostock, Germany*
*Corresponding author: e.goulielmakis@uni-rostock.de*

**We demonstrate efficient generation of coherent super-octave pulses via a single-stage spectral broadening of a Yb:KGW laser in a single, pressurized, Ne-filled, hollow-core fiber capillary. Emerging pulses spectrally spanning over more than ~1 PHz (250 nm- 1600 nm) at a dynamic range of ~60 dB, and an excellent beam quality open the door to combining Yb:KGW lasers with modern light field synthesis techniques. Compression of a fraction of the generated supercontinuum to intense (8 fs, ~2.4 cycle, ~650 µJ) pulses allows convenient use of these novel laser sources in strong-field physics and attosecond science.**

Intense pulses generated by Ti:Sapphire lasers and amplifiers have been the central element for numerous advances in modern ultrafast science and technology [1]. Spectral broadening of these pulses by their nonlinear propagation [2] in hollow-core, gas filled glass capillaries [3,4] and their compression with chirped mirrors [5] to few-cycle pulses [6] has been instrumental in strong field physics and attosecond science [7]. The development of multi-octave, light-field-synthesizers [8,9] has further allowed the compression of super-octave visible light to the single [10,11] subcycle [12] and attosecond durations [13] offering unique opportunities for exploring new realms of strong field interactions of light and matter as well as for attosecond pump-probe studies [14] of ultrafast phenomena. Over the last decade, the development of a new class of mid-infrared lasers based on the Yb:KGW technology dramatically minimized the cost and complexity of producing mJ level femtosecond laser pulses at high repetition rates. Yet, generation of supercontinuum pulses extending into the visible and UV based on these sources and their subsequent compression remains challenging.

Commercial Yb:KGW lasers, despite their high pulse energy (>1 mJ/pulse) and repetition rates (up to hundreds of kHz), are currently delivering pulses that are approximately 8-10 times longer than their Ti:Sapphire counterparts. Considering that self-phase modulation is the principal and most broadly exploited mechanism of spectral broadening of an intense laser pulse propagating in a nonlinear medium allows identifying the origin of the challenge. In self phase modulation the instantaneous frequency $\omega_{inst}(t)$ of laser pulse of an intensity profile $I(t)$, carried at a frequency $\omega_0$, resulting from the propagation of the pulse in a medium of nonlinear index $n_2$ and length $l$ is given by:

$$\omega_{inst}(t) = \omega_0 - \frac{n_2 \omega_0}{c}\frac{dI(t)}{dt}l. \quad (1)$$

Here, $c$ is the speed of light. Eq. (1) suggests that pulses from a commercial Ti:Sapphire system ($\omega_0$ ~800 nm, $\tau$ ~20 fs) and Yb:KGW ($\omega_0$ ~1030 nm, $\tau$ ~170 fs) of the same peak intensity propagating in the same nonlinear medium will undergo a dramatically different degree of spectral broadening. This primary difference stems from the steepness of the intensity envelope $\frac{dI(t)}{dt}$, which is inversely proportional to the duration of the pulse $\tau$ as well as from the carrier frequency $\omega_0$. A rough estimation based on the above parameters and Eq. (1) yields approximately a factor of ~10 higher degree of spectral broadening in favor of the Ti:Sapphire pulse.

In order for Yb:KGW lasers to serve as major tools in state-of-the-art pulse synthesis [12,13], the above-mentioned dramatic difference must be compensated. Efforts in the last decade toward this goal include pulse broadening in hollow-core, gas-filled fiber capillaries with gases of higher nonlinear index than the typically used Ne [3,6,8], such as Ar [15], Xe [16,17] as well as molecular gases [18]. Yet, the exponential increase of ionization stemming from the lower ionization energy of these media limits the highest pulse intensity that can be efficiently coupled into the waveguide, and thus constrains the attainable nonlinearity and spectral broadening in a single and compact pulse compression stage. Preliminary tests in our lab using these media (Ar and Xe) also indicated spatiotemporal instabilities of the exit beam modes, which can be attributed to the extreme nonlinearity of plasma generation from pulse to pulse. Long-length hollow-core fiber capillaries (up to 6 meters [19]) allow reducing the peak intensity of the input pulses and consequently, ionization effects, but due to the large footprint of these setups, they become less practical and prevent straightforward accommodation to existing ultrafast infrastructures. Super-octave broadening and compression of Yb:KGW laser is also possible also by cascading the processes into two or more stages using hollow-core fibers and thin glass plates [20] or self-channeling in noble gasses [21]. Yet, attaining spectral broadening that extends from the mid-infrared to the deep ultraviolet has remains beyond the reach in these systems.

A potentially ideal solution to the spectral broadening compression problem of Yb:KGW laser pulses would include: (a) single stage compression to limit complexity and maintain high beam quality; (b) straightforward and efficient compression to <3 cycle pulses with less demanding technologies (dispersive mirrors) to allow direct application to field-sensitive ultrafast experiments; (c) spectral bandwidth that allows accommodation with existing PHz scale synthesizers and infrastructures. Here, we demonstrate these possibilities.

As alluded to earlier, the replacement of Ne in hollow-core fiber compressors with gasses of lower ionization potential energy such as Ar, Kr etc. gives rise to undesirable ionization effects that are difficult to practically handle. Alternatively, one can attain the required increase in nonlinearity by a corresponding increase in the pressure of the gas in the hollow-core capillary while maintaining a high ionization potential gas such as Ne as the nonlinear medium. The general feasibility of this scheme has been previously investigated in self-channeling experiments [21]. For instance, PHz scale supercontinua driven by ~20 fs pulses in Ne [13] are typically attained at a pressure of ~2-2.5 bars. Therefore, a rough estimation based on the above discussion would suggest that reaching a similar level of spectral broadening based on an Yb:KGW source would require a Ne pressure higher than 20 bars.

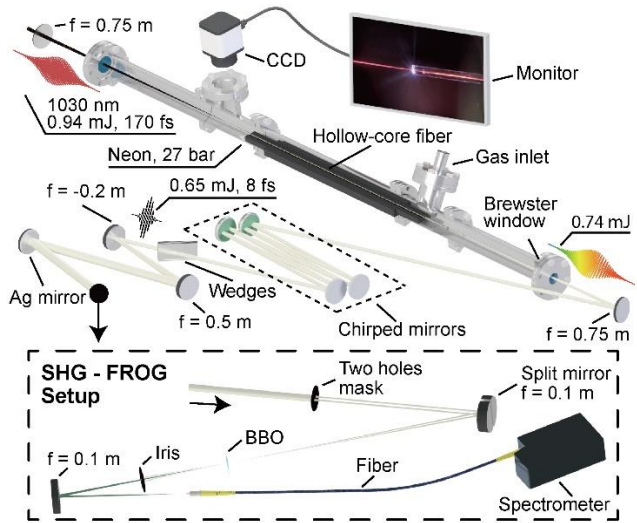

Fig. 1. Experimental setup for the generation of PHz range supercontinuum pulses from a Yb:KGW laser system and compression to few-cycle pulses. Inset shows the SHG-FROG apparatus used for the temporal characterization of the emerging laser pulses.

To explore the potential of this concept, pulses (~1 mJ, ~170 fs) generated by a commercial Yb:KGW laser system (Pharos, Light conversion) are focused ($f$ = 75 cm) into the entrance of a silica capillary (inner diameter ~250 $\mu$m, length 1.1 m) enclosed in a pressurized (up to 27 bars) neon-filled metal chamber (Fig. 1). The laser beam entered the Ne chamber through a ~2 mm thick, antireflection-coated window. A broadband and efficient transmission of the generated pulses down to UV frequencies was enabled by a UV-grade fused silica Brewster window (2 mm, 58.8°) that was installed at the exit of the host chamber. Nonlinear effects on the entrance window were minimized by placing the latter only a few centimeters away from the focusing lens, while the exit window was placed ~50 cm behind the exit of the glass capillary. The total length of the host chamber amounted to ~2.0 meters. Ordinary aluminum ISO-KF tubing typically used for manufacturing the host chamber was replaced by its stainless-steel counterparts (CF) to allow a safe operation at high gas pressure. The rigidity of the host chamber was tested at pressures beyond >30 bars of neon without any apparent damage to the entrance and exit windows.

A rough adjustment of the focus position with respect to the entrance of the silica capillary was achieved by placing the input lens on a course translation stage. Fine adjustments of the focal spot position with respect to the entrance of the capillary were enabled by setting the fiber-chamber on precision translation stages that allow the displacement of the fiber along the beam propagation axis as well as its directional adjustment with respect to the input. The entire beam path from the laser to the entrance of the hosting chamber was shielded against air fluctuations to minimize beam pointing. A CCD camera (Fig. 1) imaged the entrance of the capillary through a view port of the host chamber and allowed for a convenient inspection of the beam coupling into the capillary. The latter turned out to be particularly useful for a well-controlled, reproducible, and rapid day-to-day optimization of the system as well as the prevention of accidental laser damage to the capillary entrance during alignment.

The spectra of the emerging supercontinuum pulses were recorded at the exit of the host chamber upon reflection off a specular reflectance standard (Ocean Optics). To allow a broadband spectral measurement, we combined two spectrometers for 200 nm-1100 nm (HR4000, Ocean Optics) and 900 nm-1700 nm (NIRQuest, Ocean Optics) using a bifurcated optical fiber. The spectral intensity calibration over the range (250 nm-1300 nm) was conducted using a precalibrated tungsten-deuterium source (DH-3P-CAL, Ocean Optics).

Fig. 2(a) shows spectra recorded at representative pressure settings of Ne in the chamber. At 5 bars of Ne [green line, Fig. 2(a)], the spectral width of the generated supercontinuum spectra is nearly duplicated in comparison to the original pulse spectrum [gray line, Fig. 2(a)]. The bandwidth is further increased at higher pressure [cyan line, Fig. 2(a)] and reaches a bandwidth that spans the range from ~250 nm to 1600 nm, that is ~1 PHz [blue line, Fig. 2(a)] at a pressure of ~27 bars and over a dynamic range of ~60 dB. Importantly, we note the coherent generation of frequencies in the UV/VUV whose importance in attosecond synthesis techniques has been previously highlighted [13]. At this pressure setting, the transmission through the total system (chamber and capillary) was ~80% corresponding to an energy of ~700 $\mu$J per pulse at its exit. A conspicuous, broad spectral peak at ~300 nm [blue line, Fig. 2(a)] can be attributed to third harmonic generated by the propagation of input pulses in the gas medium, and which is also spectrally broadened upon copropagation with the supercontinuum pulse. Such features have been previously investigated in the propagation of few cycle pulses in pressurized media [22].

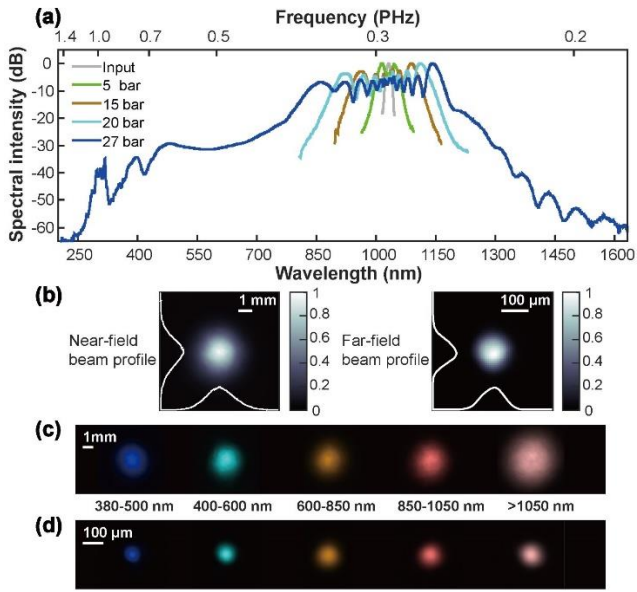

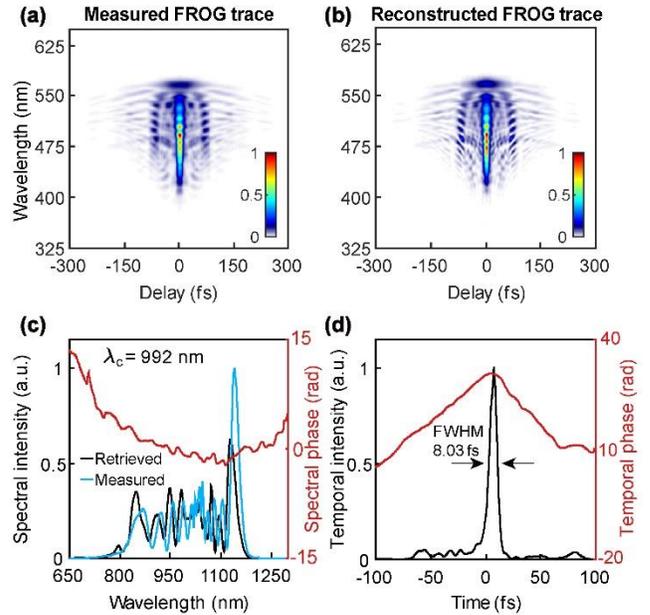

Fig. 2. (a) Supercontinuum spectra for various pressure setting of Ne in the hollow-core capillary. (b) Near- and far-field beam profile of the supercontinuum pulses. (c) Near- and (d) far-field beam profiles of the generated pulses at different spectral bands.

Fig. 3. Few-cycle pulse measurement. (a) Measured SHG-FROG trace. (b) Reconstructed SHG-FROG trace (c) Measured spectrum (cyan) and retrieved spectrum (black) and spectral phase (red). $\lambda_c$ denotes the centroid wavelength of the reconstructed spectrum. (d) Temporal intensity profile (black) and corresponding temporal phase (red) of the retrieved pulse. The evaluated pulse duration at the FWHM is ~8 fs corresponding to ~2.4 cycles of the carrier frequency.

To investigate the spatial properties of the generated supercontinuum pulses, we used beam profiler to record both near- and far-field beam profiles of the supercontinuum pulses [Fig. 2(b)]. Measurements at different spectral ranges [as indicated in Fig. 2(c) and Fig. 2(d)] suggest an excellent beam quality for both near- and far- field beam profiles reflect the well-established capability offered by hollow-core fiber compressors to deliver pulses of high beam quality.

To assess the direct compressibility of the generated supercontinuum pulses and explore the possibility to generate few-cycle pulses with readily available pulse compression tools, we used dispersive mirrors designed for the range 650 nm-1350 nm (PC147, UltraFast Innovations). This spectral range corresponds to approximately the top 25 dB of the generated supercontinuum, that is, spectra components that can significantly contribute to the pulse shortening. Indeed, the Fourier limited duration of a pulse within this spectral range corresponds to approximately ~6 fs. The total transmission of the supercontinuum pulses through the dispersive mirror compressor was ~88% resulting in a total energy of ~650 µJ per pulse available to experiments. Optimal temporal compression was achieved by using 8 reflections (-60 fs$^2$ per reflection). Fine-tuning of the pulse chirp was achieved by a pair of thin fused-silica wedges installed in the beam path. A home-built, all-reflective (aluminum mirrors) second harmonic, frequency resolved optical gating apparatus [23] (SHG-FROG) was used to characterize pulses at the compressor exit. To enable a broadband phase matching of the second harmonic signal, a thin BBO crystal (~5 µm) mounted on a fused silica substrate was used.

Fig. 3(a) shows a representative SHG-FROG trace recorded with our apparatus. The reconstructed trace, based on a projections-based algorithm [24] is shown in Fig. 3(b) and suggests a fair reconstruction fidelity. The latter is further verified by a low FROG error (0.0036 for a 1024 × 1024 matrix) and a good agreement between the pulse spectrum [cyan line, Fig. 3(c)] measured at the exit of the chirp mirror compressor (650 nm-1350 nm) and its retrieval [black line, Fig. 3(c)]. The duration of the retrieved pulse was ~8 fs, measured at the full width at half maximum of its intensity profile. The centroid carrier frequency of the pulse was ~0.3 PHz corresponding to a wavelength of ~992 nm. This implies that the pulse contains ~2.4 field-cycles within the FWHM of its intensity envelope. A Fourier-limited duration ~6.2 fs corresponding to ~1.9 cycles of the pulse shall also become attainable using a customized chirped mirror set.

In conclusion, we generate supercontinua spanning over more than 1 PHz around a central wavelength of 992 nm by nonlinear broadening of Yb:KGW pulses (~1 mJ, 6 kHz, 1030 nm) in a Ne-filled, pressurized hollow-core fiber capillary. A transmission efficiency >75% and a high beam quality over the entire spectral range are also demonstrated. We compress a part of the supercontinuum (~650 nm to 1200 nm) into few-cycle pulses (~8 fs) whose intensity envelope is confined to ~2.4 cycles of their carrier frequency. The PHz range supercontinua extending to the ultraviolet part of the spectrum open the door to integrating Yb:KGW laser technologies with state-of-the-art synthesizers. Extension of the new supercontinuum source towards the mid-infrared (1000 nm-1500 nm) will add a new dimension to light field synthesis based on single-stage supercontinuum sources.

**Funding.** "Funded by the Deutsche Forschungsgemeinschaft (DFG, German Research Foundation) - SFB 1477 "Light-Matter Interactions at Interfaces", project number 441234705".

**Disclosures.** The authors declare no conflicts of interest.

**Data availability.** Data underlying the results presented in this paper are not publicly available at this time but may be obtained from the authors upon reasonable request.

## References

1. T. Brabec and F. Krausz, "Intense few-cycle laser fields: Frontiers of nonlinear optics," Rev Mod Phys 72, 545–591 (2000).
2. A. M. Zheltikov, "Let there be white light: supercontinuum generation by ultrashort laser pulses," Physics-Uspekhi 49, 605 (2006).
3. M. Nisoli, S. de Silvestri, and O. Svelto, "Generation of high energy 10 fs pulses by a new pulse compression technique," Appl Phys Lett 68, 2793–2795 (1996).
4. C. Vozzi, M. Nisoli, G. Sansone, S. Stagira, and S. de Silvestri, "Optimal spectral broadening in hollow-fiber compressor systems," Applied Physics B 80, 285–289 (2005).
5. R. Szipöcs, K. Ferencz, C. Spielmann, and F. Krausz, "Chirped multilayer coatings for broadband dispersion control in femtosecond lasers," Opt Lett 19, 201–203 (1994).
6. M. Nisoli, S. de Silvestri, O. Svelto, R. Szipöcs, K. Ferencz, Ch. Spielmann, S. Sartania, and F. Krausz, "Compression of high-energy laser pulses below 5 fs," Opt Lett 22, 522–524 (1997).
7. F. Krausz and M. Ivanov, "Attosecond physics," Rev Mod Phys 81, 163–234 (2009).
8. M. Th. Hassan, A. Wirth, I. Grguraš, A. Moulet, T. T. Luu, J. Gagnon, V. Pervak, and E. Goulielmakis, "Invited Article: Attosecond photonics: Synthesis and control of light transients," Review of Scientific Instruments 83, 111301 (2012).
9. C. Manzoni, O. D. Mücke, G. Cirmi, S. Fang, J. Moses, S.-W. Huang, K.-H. Hong, G. Cerullo, and F. X. Kärtner, "Coherent pulse synthesis: towards sub-cycle optical waveforms," Laser Photon Rev 9, 129–171 (2015).
10. A. L. Cavalieri, E. Goulielmakis, B. Horvath, W. Helml, M. Schultze, M. Fieß, V. Pervak, L. Veisz, V. S. Yakovlev, M. Uiberacker, A. Apolonski, F. Krausz, and R. Kienberger, "Intense 1.5-cycle near infrared laser waveforms and their use for the generation of ultra-broadband soft-x-ray harmonic continua," New J Phys 9, 242–242 (2007).
11. E. Goulielmakis, V. S. Yakovlev, A. L. Cavalieri, M. Uiberacker, V. Pervak, A. Apolonski, R. Kienberger, U. Kleineberg, and F. Krausz, "Attosecond Control and Measurement: Lightwave Electronics," Science (1979) 317, 769–775 (2007).
12. A. Wirth, M. Th. Hassan, I. Grguraš, J. Gagnon, A. Moulet, T. T. Luu, S. Pabst, R. Santra, Z. A. Alahmed, A. M. Azzeer, V. S. Yakovlev, V. Pervak, F. Krausz, and E. Goulielmakis, "Synthesized Light Transients," Science (1979) 334, 195 LP – 200 (2011).
13. M. T. Hassan, T. T. Luu, A. Moulet, O. Raskazovskaya, P. Zhokhov, M. Garg, N. Karpowicz, A. M. Zheltikov, V. Pervak, F. Krausz, and E. Goulielmakis, "Optical attosecond pulses and tracking the nonlinear response of bound electrons," Nature 530, 66–70 (2016).
14. A. Moulet, J. B. Bertrand, T. Klostermann, A. Guggenmos, N. Karpowicz, and E. Goulielmakis, "Soft x-ray excitonics," Science (1979) 357, (2017).
15. X. Guo, S. Tokita, K. Yoshii, H. Nishioka, and J. Kawanaka, "Generation of 300 nm bandwidth 0.5 mJ pulses near 1 $\mu$m in a single stage gas filled hollow core fiber," Opt Express 25, 21171–21179 (2017).
16. S.-Z. A. Lo, L. Wang, and Z.-H. Loh, "Pulse propagation in hollow-core fiber at high-pressure regime: application to compression of tens of $\mu$J pulses and determination of nonlinear refractive index of xenon at 1.03 $\mu$m," Appl Opt 57, 4659–4664 (2018).
17. J. E. Beetar, F. Rivas, S. Gholam-Mirzaei, Y. Liu, and M. Chini, "Hollow-core fiber compression of a commercial Yb:KGW laser amplifier," Journal of the Optical Society of America B 36, A33–A37 (2019).
18. J. E. Beetar, M. Nrisimhamurty, T.-C. Truong, G. C. Nagar, Y. Liu, J. Nesper, O. Suarez, F. Rivas, Y. Wu, B. Shim, and M. Chini, "Multioctave supercontinuum generation and frequency conversion based on rotational nonlinearity," Sci Adv 6, eabb5375 (2022).
19. Y.-G. Jeong, R. Piccoli, D. Ferachou, V. Cardin, M. Chini, S. Hädrich, J. Limpert, R. Morandotti, F. Légaré, B. E. Schmidt, and L. Razzari, "Direct compression of 170-fs 50-cycle pulses down to 1.5 cycles with 70% transmission," Sci Rep 8, 11794 (2018).
20. C.-H. Lu, W.-H. Wu, S.-H. Kuo, J.-Y. Guo, M.-C. Chen, S.-D. Yang, and A. H. Kung, "Greater than 50 times compression of 1030 nm Yb:KGW laser pulses to single-cycle duration," Opt Express 27, 15638–15648 (2019).
21. E. Goulielmakis, S. Koehler, B. Reiter, M. Schultze, A. J. Verhoef, E. E. Serebryannikov, A. M. Zheltikov, and F. Krausz, "Ultrabroadband, coherent light source based on self-channeling of few-cycle pulses in helium," Opt Lett 33, 1407–1409 (2008).
22. U. Graf, M. Fieß, M. Schultze, R. Kienberger, F. Krausz, and E. Goulielmakis, "Intense few-cycle light pulses in the deep ultraviolet," Opt Express 16, 18956–18963 (2008).
23. K. W. DeLong, R. Trebino, J. Hunter, and W. E. White, "Frequency-resolved optical gating with the use of second-harmonic generation," Journal of the Optical Society of America B 11, 2206–2215 (1994).
24. K. W. DeLong, D. N. Fittinghoff, R. Trebino, B. Kohler, and K. Wilson, "Pulse retrieval in frequency-resolved optical gating based on the method of generalized projections," Opt Lett 19, 2152–2154 (1994).